\documentclass[twoside]{mhd}
\usepackage{graphicx}
\usepackage{bm}
\mhdhead{40}{1}{1}

\title{\uppercase{Schwabe, Gleissberg, Suess-de Vries: Towards a consistent
model of planetary synchronization of solar cycles}}

\author{F.~Stefani\inst{1*}, A.~Giesecke\inst{1}, M. Seilmayer\inst{1}, 
R. Stepanov\inst{2}, T.~Weier\inst{1}}

\institute{Helmholtz-Zentrum Dresden -- Rossendorf, Bautzner Landstr. 400, 
01328 Dresden, Germany
\and
Institute of Continuous Media Mechanics, Acad. Korolyov str. 1, 614013 Perm, Russia
} 

\begin{document}
\maketitle

\noindent
\textbf{Abstract:}
Aiming at a consistent planetary synchronization model 
of both short-term and long-term solar cycles, we start with an analysis 
of Schove's historical data of cycle maxima. 
Their deviations (residuals) from the average cycle duration 
of 11.07 years show a high degree of regularity, comprising a dominant 
200-year period (Suess-de Vries cycle), and a few periods 
around 100 years (Gleissberg cycle).
Encouraged by their robustness, 
we support previous forecasts of an upcoming grand 
minimum in the 21st century. To explain 
the long-term cycles, 
we enhance our tidally synchronized solar dynamo model 
by a modulation of the field storage capacity of the tachocline with 
the orbital angular momentum of the Sun, which is dominated by the
19.86-year periodicity of the Jupiter-Saturn synodes.
This modulation of the 22.14 years Hale cycle 
leads to a 193-year beat 
period of dynamo activity which is indeed close to the 
Suess-de Vries cycle. For stronger dynamo modulation, 
the model produces additional peaks at typical 
Gleissberg frequencies, which seem to be explainable by the
non-linearities of the basic beat process, leading
to a bi-modality of the Schwabe cycle. However, 
a complementary role of 
beat periods between the Schwabe cycle  
and the 
Jupiter-Uranus/Neptune synodic cycles 
cannot be completely excluded.


\section{Introduction}
\label{sec:intro}

Solar activity is governed by the 11-year Schwabe cycle
(or 22-year Hale cycle), and a few long-term cycles superposed
on it \cite{Charbonneau2010,Hathaway2010}. Among those, the Gleissberg cycle (90 years) 
and the Suess-de Vries cycle (200 years) figure most 
prominently, while the Hallstatt cycle 
(2300 years) may play a ``super-modulating'' role \cite{Beer2018}. 
Recent solar dynamo models
have been successful in understanding both the typical time scale of the 
Schwabe cycle as well as the shape of the butterfly diagram of sunspots.
In the framework of non-linear dynamo models, the disparity 
between short- and long-term cycles was explained as a 
consequence of the small magnetic Prandtl number in the tachocline 
region \cite{Tobias1996}. 

This being said, some features of the solar cycles
leave us with the nagging feeling that conventional dynamo 
models might not be the end of the story. As a case in point, 
Dicke's ratio \cite{Dicke1978} 
of the mean square of the residuals (i.e. the distances between the 
actual minima and the hypothetical minima of a perfect 
11.07-year cycle) to the mean square of the differences 
between two consecutive residuals, points decisively to a clocked 
process, in stark contrast to a random walk process \cite{Stefani2019}. 
As for the long-term cycles, 
it is foremost the sharpness of the 
200-year Suess-de Vries cycle  which 
leaves us unconvinced about the 
skill of present dynamo models for explaining it. 
Even if one denies Dicke's 
question ``Is there chronometer 
hidden {\it deep in the Sun}?'' \cite{Dicke1978}, it is the 
counter-question concerning a possible
{\it external} clock for the solar dynamo 
which leads us, 
rather inevitably, into the field of planetary synchronization.

Despite the long history of planetary synchronization models, 
going back to early speculations of Wolf \cite{Wolf1859}, 
there has always been some 
air of  ``astrology'' hanging about them. Distinguishing 
between models based on tidal forcing and models based on 
spin-orbit coupling, there is indeed good reason for profound 
skepticism towards both of them.
Tidal forcing models can easily be ridiculed by the tiny
acceleration of $\sim 10^{-10}$\,ms$^{-2}$ as exerted 
by planets \cite{Callebaut2012},
leading to a negligible tidal height of not 
more than 1\,mm.
Spin-orbit models \cite{Zaqa1997,Juckett2000} have likewise been criticized 
\cite{Shirley2006} 
for not being able to conclusively
explain how any {\it internal} differential motion could be produced  
from the  free-fall motion of the Sun around
the solar system's barycenter (SSB), however impressive the  
amplitude of that motion (around 1 solar diameter)  
and its speed (until 15\,m/s) may ever appear.    

Even if recognizing the seriousness of such objections, 
some more specific considerations seem appropriate.
As for the tidal force, one should note that the
typical tidal height, as produced  by a planet of mass $m$ at 
distance $d$ from the Sun, 
$h_{\rm tidal} =
G m R^2_{\rm tacho}/(g_{\rm tacho} d^3)= O(1$\,mm)  
translates  - via virial theorem - into a non-negligible velocity
of $v \sim (2 g_{\rm tacho} h_{\rm tidal})^{1/2} \approx 1$\,m/s, 
when employing the huge gravity at the tachocline of 
$g_{\rm tacho} \approx 500$\,m/s$^{2}$.
Likewise, for the spin-orbit model it has been argued 
\cite{Wilson2008,Sharp2013} that some 0.1 per 
cent of the typical orbital angular momentum
variation of the Sun ($5\times 10^{40}$\,Nms) 
might well be transferred into internal differential motion, 
which would amount to a velocity scale of 4\,m/s when applied
only to the 2 per cent of the total solar mass as
concentrated in the convection zone. 
Again, velocities of that scale could definitely be dynamo relevant, 
remembering  a similar scale of 10\,m/s for the meridional circulation 
\cite{Charbonneau2010}. 
Partially related to the distinction between {\it tidal} versus {\it spin-orbit} 
models, planetary forcing models can further be classified
into models of hard {\it synchronization} of the basic Schwabe cycle
(for example, with the 11.07 years spring tide period of the tidally 
dominant Venus-Earth-Jupiter system \cite{Stefani2019,Hung2007,Wilson2013,Okhlopkov2014,Stefani2016,Stefani2018}) 
and models of  
soft {\it modulations} of this Schwabe cycle, with
main focus on the  Gleissberg, 
Suess-de Vries and Hallstatt cycle
\cite{Jose1965,Fairbridge1987,Charvatova1997,Landscheidt1999,Wolffpatrone2010,Abreu2012,McCracken2014,Cionco2015,Scafetta2016}. 

Instead of linking the periods of Sun's long-term cycles to
corresponding periods of planetary influences, 
we pursue here another concept, in which long-term 
cycles emerge as {\it beat periods} between the basic 
Hale/Schwabe cycle with the typical synodic periods of Jupiter 
with other Jovian planets.
The beat period of 193 years, as resulting from the 22.14-year Hale cycle 
and the 19.86-year synodic cycle of Jupiter and Saturn, 
has been noticed by several authors \cite{Wilson2013,Solheim2013}.
Likewise, one may wonder if the beat periods 55.8 years and 82.7 years, arising 
between the 11.07 Schwabe 
cycle and the Jupiter-Uranus synode (13.81 years) and  the Jupiter-Neptune 
synode (12.78 years), respectively,  could 
somehow be related to the Gleissberg cycle.

The aim of this paper is to corroborate how such beat periods 
actually emerge in our specific solar dynamo model \cite{Stefani2019}, which had 
already demonstrated synchronization of the Schwabe cycle with the 11.07 years
tidal period, based on the resonance of the intrinsic helicity oscillations 
of the current-driven $m=1$ Tayler instability with the $m=2$ tidal forcing
\cite{Weber2015,Stepanov2019}. Before entering this topic, 
we reconstruct the typical Gleissberg and Suess-de Vries periods from
the long series of solar cycle maxima data as bequeathed to us by 
Schove \cite{Schove1955,Schove1979,Schove1983}. The paper will close with a summary and  
a short discussion of open issues.

\section{Spectral analyses of cycle maxima}

In a meticulous effort over three decades \cite{Schove1955,Schove1979,Schove1983}, 
Schove had tried to identify the minima and maxima of the solar cycle for 
nearly two and a half millennia, relying strongly on 
historical {\it aurora borealis} observations. 
Although Schove's time series are 
considered by some researchers as ``archaic'' \cite{Usoskin2017}, they are 
still hard to replace when it comes to the very dating of individual 
maxima and minima for early times 
(hopefully, a careful analysis of existing $^{14}$C or $^{10}$Be data, 
e.g. \cite{Berggren2009}, may once allow to verify, or falsify, 
Schove's time series, at least for A.D. 1400 onward).
Based on Schove's {\it minima data}, in \cite{Stefani2019} we had 
computed Dicke's ratio \cite{Dicke1978} to argue in favour of the solar 
cycle as being a clocked, rather than a random 
walk process. In Fig.~1a we show now the sequence 
of {\it maxima data}\footnote{One reason for 
using the maxima data (which are commonly considered less accurate than the 
minima data) is the annoying fact that the list of Schove's minima 
- Appendix A in his most recent and comprehensive publication 
\cite{Schove1983} - is spoiled by wrong data between A.D. 511 - 1493; 
those were falsely copied from the table of the corresponding maxima.}
after having subtracted
two different linear functions with an 11.07-year and an 11.11-year trend, 
respectively. 
Obviously, 
the residuals from the 11.07-year trend form a 
rather horizontal band, while the residuals from the 11.11-year 
trend still exhibit an unresolved downward inclination. 
This slight, but important, 
difference disproves the often heard objection against Schove's data 
(e.g. \cite{Usoskin2017}, taking a 
loose remark on page 131 of \cite{Schove1955} too literally)
as being biased by a strict constraint of ``9 maxima in 100 years''. 
The fact that the 11.07-year trend appears naturally 
from Schove's data, without ever being enforced by him, speaks 
strongly in favour of the validity of that cycle period.
Another interesting feature is the 
bi-modality of the histogram of the cycle durations (Fig. 1 c), 
with two maxima at around 10 and 12 years 
flanking the so-called Wilson gap \cite{Hathaway2010}, which 
will play an important role in our further analysis.

\begin{figure}[t]
  \centering
  \includegraphics[width=0.99\textwidth]{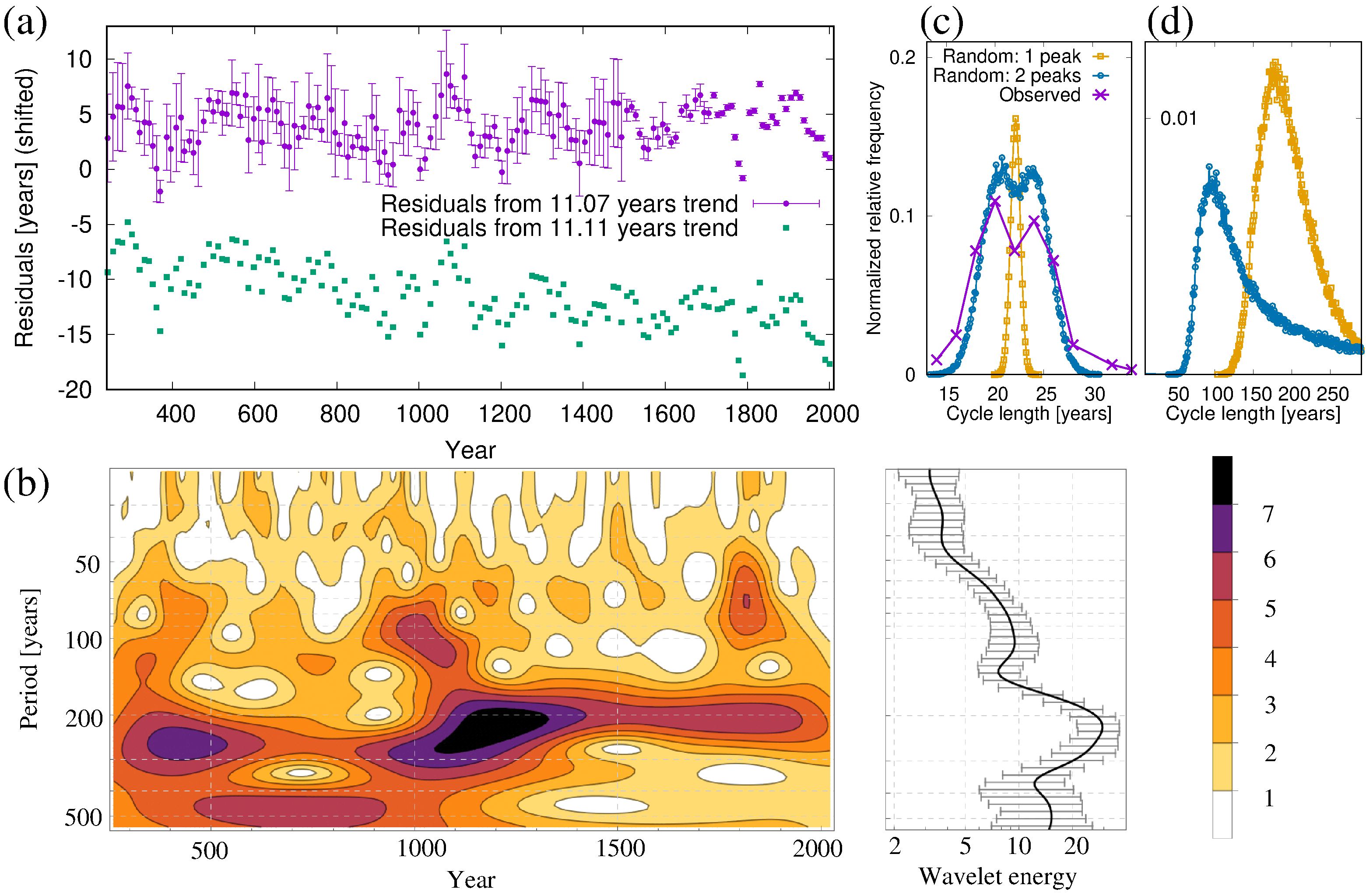}
  \caption{(a) Residuals of the cycle maxima \cite{Schove1983} from two
  trends with 11.07-year (including error bars) 
  and 11.11-year period (differently shifted). 
  (b) Wavelet analysis of the 
  data of (a). Error bars of the wavelet energy indicate 95 per cent 
  two-sided confidence intervals. (c) Purple - Histogram of the (doubled) 
  cycle lengths according to (a). Orange - Histogram (reduced by factor 5) 
  of 10$^5$ random numbers $T_{\rm ran}$
  centered narrowly around 22.14 years. Blue - Histogram of 10$^5$ random 
  numbers $T_{\rm ran}$ bi-modally centered around 19.86 and 24.42 years 
  (similar as observed data). 
  (d) Histogram 
  of arising beat periods $19.86\; T_{\rm ran}/|T_{\rm ran}-19.86|$ 
  using the random numbers $T_{\rm ran}$ from (c). 
  While the narrowly centered $T_{\rm ran}$ (orange)
  produce a dominant beat period around 180 years, the 
  bi-modally centered $T_{\rm ran}$ (blue) produce a beat period 
  around 90 years.
  }
  \label{fig:1}
\end{figure}

While long-term solar cycles are usually inferred from various
continuous data sets (e.g., $^{10}$Be or $^{14}$C isotopes 
\cite{McCracken2014,Muscheler2007} 
or paleoclimate data \cite{Luedecke2015}), 
quite similar cycle periods can also be identified by analyzing the 
time series of the residuals of the cycle maxima or minima.
The ``(O-C) residuals'', as computed in \cite{Richards2009} 
for data between A.D. 1610 - 1996, have revealed peaks at a 
Suess-de Vries period of 188 years, a Gleissberg cycle of 
87 years, and an additional (unnamed) 
40 years cycle. Here, we take advantage of Schove's much longer
data-set of cycle maxima, 
complemented by Hathaway's data \cite{Hathaway2010} 
for the later years of the total 
interval A.D. 242 - 2000. The starting year
A.D. 242  has been chosen since the value before that date
is the first one that is missing in \cite{Schove1983}. For frequency analysis, 
we utilize first a wavelet transform 
(Fig.~1b, using the Morlet wavelet with resolution parameter $2 \pi$ and 
a gaped wavelet algorithm for solving the edge problem), and additionally a
generalized Lomb-Scargle method (Fig.~2) 
which accounts for different prior 
uncertainties for individual data points. 
Specifically, our data and their uncertainties 
are compiled as follows: for the early years 
A.D. 241 - 1493 they are taken from Appendix B of Schove's most recent 
and comprehensive 
publication  \cite{Schove1983}. As uncertainties, we use 
the comparably large values from the 
older publication \cite{Schove1955}. Between A.D. 1506 - 1700, we 
use the maxima data and their uncertainties from Table 2 of 
\cite{Schove1983}. From
A.D. 1718 onward, we opted for a somewhat 
optimistic uncertainty of 0.25 years, not least 
in order to give those recent data a higher statistical weight 
compared to the older ones.  The data 
from  A.D. 1718  - 1761 are taken again from \cite{Schove1983}, whilst 
all subsequent maxima are taken from the more modern table of 
\cite{Hathaway2010} (they mainly coincide with the data
of \cite{Schove1983}).

\begin{figure}[t]
  \centering
  \includegraphics[width=0.99\textwidth]{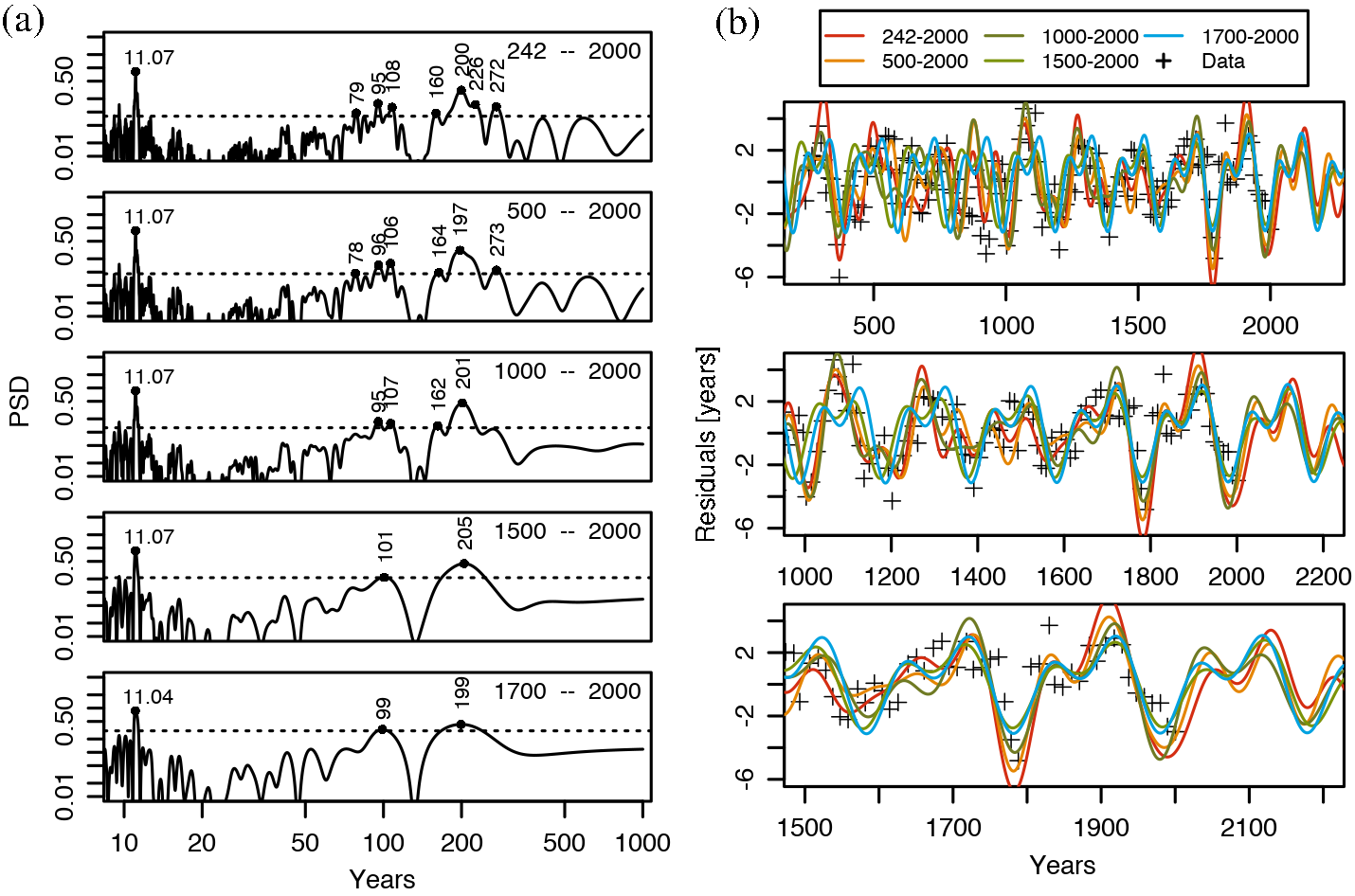}
  \caption{(a) Lomb-Scargle periodogram for the maxima residuals, 
  when taking into account data (and their uncertainties) from five 
  different time intervals. The dashed horizontal line in each sub-panel
  indicates a 25 per cent false alarm probability. (b) Complete picture
  and two zooms of the residuals, showing  also five different approximations 
  with those harmonics that are statistically 
  significant for the respective fitting intervals, as indicated 
  by full circles in (a).  Cycles longer (shorter) than 11.07 years correspond 
  to a positve (negative) slope of the curves.
  }
  \label{fig:2}
\end{figure}

With view on the
decreasing reliableness of Schove's data before A.D. 1600, say, we compute 
the generalized Lomb-Scargle periodograms (Fig.~2a) for different underlying 
time intervals. Apart from a general broadening
of the peaks (and some minor shifts), when going over to shorter (i.e., later) 
time intervals, we observe quite robust features such as a 
dominant Suess-de Vries cycle around 200 years 
(which also dominates the wavelet spectrum, see Fig. 1b) 
and one or a few 
Gleissberg-type cycles around 100 years, 
in good agreement with \cite{Solheim2013,Richards2009}.
The generalized Lomb-Scargle method provides us also with a significance 
level (here: 25 per cent false alarm probability, indicated by the dashed 
lines in Fig.~2a)
to identify the most significant peaks for each underlying time interval.
The various data reconstructions with the corresponding sets of
significant harmonics (as indicated by the full circles 
in Fig.~2a) are shown in Fig.~2b.
Interestingly, the extrapolations of the various fit curves
show a rather consistent tendency towards 
longer solar cycles throughout the 21st century.
With regard to the inverse relationship between the cycle length
and the amplitude of the (following) cycle \cite{Hathaway2010},
we support the forecasts of a  new grand minimum that 
were already made by various authors 
\cite{Landscheidt1999,Solheim2013,Richards2009}.

\section{A synchronized and modulated dynamo model}

We have seen that the period of the dominant Suess-de Vries
cycle as inferred from Schove's maxima data is very close to 200 years, 
which is highly consistent with previous results based on 
$^{10}$Be and $^{14}$C data \cite{Muscheler2007}, and various climate 
related data \cite{Luedecke2015}. 
The robustness and relative sharpness of that peak suggests
a link to planetary forcings with equal or similar periods, 
as discussed by many authors \cite{Jose1965,Fairbridge1987,Charvatova1997,Landscheidt1999,Wolffpatrone2010,Abreu2012,McCracken2014,Cionco2015,Scafetta2016}. Another explanation, which also brings us
close to the 200 years cycle, relies on the beat period of 193 
years \cite{Wilson2013,Solheim2013}
that arises from the interplay of the 22.14-year Hale cycle and 
the 19.86-year synodic cycle of Jupiter and Saturn (which produces, 
according to Fig.~3, the dominant 
component of the solar motion around the SSB). A similar 
beat mechanism between the 11.07-year Schwabe cycle and 
the 13.81-year Jupiter-Uranus 
and/or the 12.78-year Jupiter-Neptune synode may likewise be considered 
a candidate for producing 
Gleissberg-type periods of 55.8 years and 82.7 years, 
respectively.

\begin{figure}[t]
  \centering
  \includegraphics[width=0.88\textwidth]{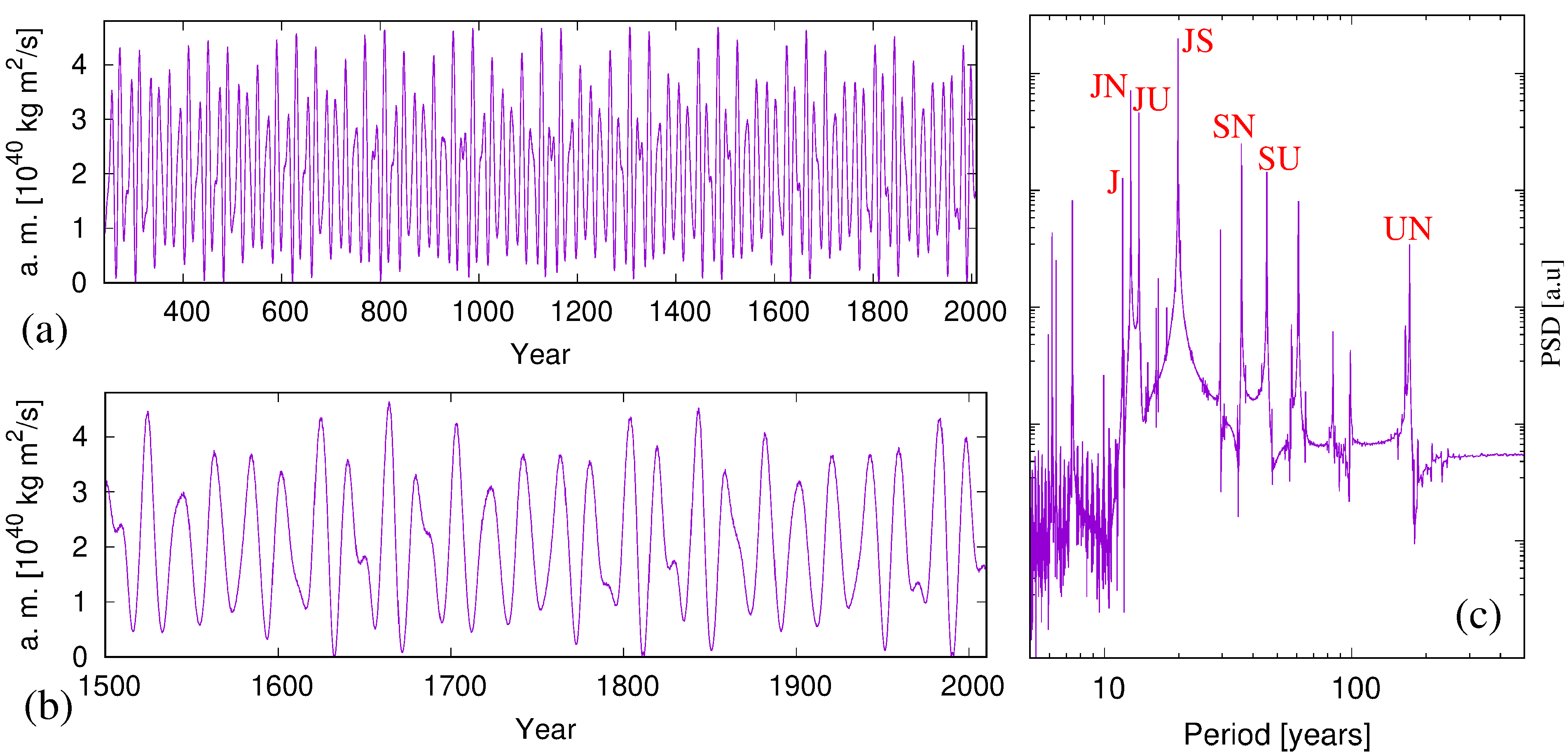}
  \caption{(a) Time series of the orbital angular momentum (a.m.) 
  of the Sun around the SSB in the interval A.D. 240-2001, and (b) zoom 
  thereof for A.D. 1500-2001, based on the DE431 ephemerides  \cite{Folkner2014}. 
  (c) PSD of the angular momentum for the long interval 13199 B.C.-A.D. 17000, with 
  some individual peaks attributed to planetary synodes (cf. \cite{Scafetta2016}): 
  JN: Jupiter-Neptune (12.78 years), JU: Jupiter-Uranus (13.95 years), 
  JS: Jupiter-Saturn (19.86 years), 
  SN: Saturn-Neptune (35.87 years), SU: Saturn-Uranus (45.36 years),
  UN: Uranus-Neptune (171.39 years). J indicates the 11.86 years period 
  of Jupiter.
  }
  \label{fig:3}
\end{figure}

To corroborate this idea, we extend the dynamo model of 
\cite{Stefani2019},
in which the Hale cycle was produced by synchronizing a 
conventional $\alpha-\Omega$ dynamo 
with an additional 11.07-year oscillation of the $\alpha$ effect.
This oscillation of $\alpha$ 
(related to the helicity of the $m=1$ Tayler 
instability or, alternatively, an 
$m=1$ magneto-Rossby wave \cite{Dikpati2017,Zaqa2018}) 
was, in turn, assumed to be resonantly excited by an 
$m=2$ planetary tidal forcing. As in \cite{Stefani2019}, we use 
the equation system
\begin{eqnarray} 
  \frac{{\partial} B(\theta,t)}{{\partial} t} &=& \omega(\theta,t) \frac{\partial A(\theta,t)}{\partial \theta} 
  + \frac{\partial^2 B(\theta,t)}{\partial \theta^2} -\kappa(t) B^3(\theta,t) ,\\
    \frac{{\partial} A(\theta,t) }{{\partial} t} &=& \alpha(\theta,t) B(\theta,t) 
    + \frac{\partial^2 A(\theta,t)}{\partial \theta^2}  
    \label{system_tayler}
   \end{eqnarray}
for the vector potential $A(\theta,t)$  of the 
poloidal field at co-latitude $\theta$ 
and time $t$, and the toroidal field
$B(\theta,t)$. 

According to \cite{Charbonneau2010} we employ
a $\theta$-dependence of the $\omega$-effect
in the form 
\begin{eqnarray}
\omega(\theta)&=&\omega_0 (1-0.939-0.136 \cos^2(\theta)-0.1457 \cos^4(\theta) )\sin(\theta) \nonumber
\end{eqnarray}
with a plausible value 
$\omega_0=10000$. The helical source term $\alpha$ comprises, first, a non-periodic 
part 
\begin{eqnarray}
\alpha^c(\theta,t)&=&\alpha^c_0(1+\xi(t))  \sin(2 \theta)/{(1+q^c_{\alpha} B^2(\theta,t))}, \nonumber
\end{eqnarray}
with a constant $\alpha^c_0$ and a noise term $\xi(t)$, and second,
a periodic part 
\begin{eqnarray}
\alpha^p(\theta,t)&=&\alpha^p_0 \sin(2 \pi t/11.07) B^2(\theta,t)/(1+q^p_{\alpha} B^4(\theta,t)) S(\theta), \nonumber
\end{eqnarray}
where $S(\theta)$ is a hemispherically  asymmetric (and slightly smoothed) term
that is non-zero only for $55^{\circ}<\theta<125^{\circ}$.
The noise $\xi(t)$, defined by the correlator
\begin{eqnarray}
\langle \xi(t) \xi(t+t_1) \rangle = D^2 (1-|t_1|/t_{\rm corr}) \nonumber
\Theta(1-|t_1|/t_{\rm corr}),
\end{eqnarray}
is numerically realized by
random numbers with variance $D^2$ which are held constant 
over a correlation time $t_{\rm corr}$. For more details of the numerical 
model, see \cite{Stefani2019}.

The term $\kappa(t) B^3(\theta,t)$ 
had been included to account 
for losses owing to magnetic buoyancy.  While we 
openly admit that the necessary spin-orbit type coupling mechanism
of the orbital angular momentum of the Sun around the SSB 
into some dynamo 
relevant parameters remains an open question (for ideas, 
see \cite{Zaqa1997,Wilson2008,Sharp2013,Palus2000}) 
we employ in the following a modulation of the parameter
$\kappa$  with the time series 
of the angular momentum. 
Since $\kappa$  is related to 
the adiabaticity in the tachocline which is, in turn, a very sensitive 
parameter \cite{Abreu2012}, its modification by some sort
of spin-orbit coupling seems, at least, not completely unrealistic.

What is now the effect of this modulation of $\kappa$ on the 
dynamo process?
Figure 4 illustrates paradigmatic solutions of Eqs. (1-2) with increasing
complexity. First, Fig.~4a shows
a conventional $\alpha-\Omega$ dynamo, without any synchronization, i.e., 
with $\alpha^p_0=0$. This specific dynamo happens to produce a {\it quadrupole} field
with an oscillation period of 24.39 years which is close, but 
not identical, to the Hale cycle (see the PSD in Fig.~4f). When 
adding to this dynamo an oscillatory $\alpha$-term 
with $\alpha^p_0=50$, we obtain the clear {\it dipole} 
configuration of Fig.~4b, synchronized now to the precise Hale 
period of 22.14 years.

\begin{figure}[t!]
  \centering
  \includegraphics[width=0.97\textwidth]{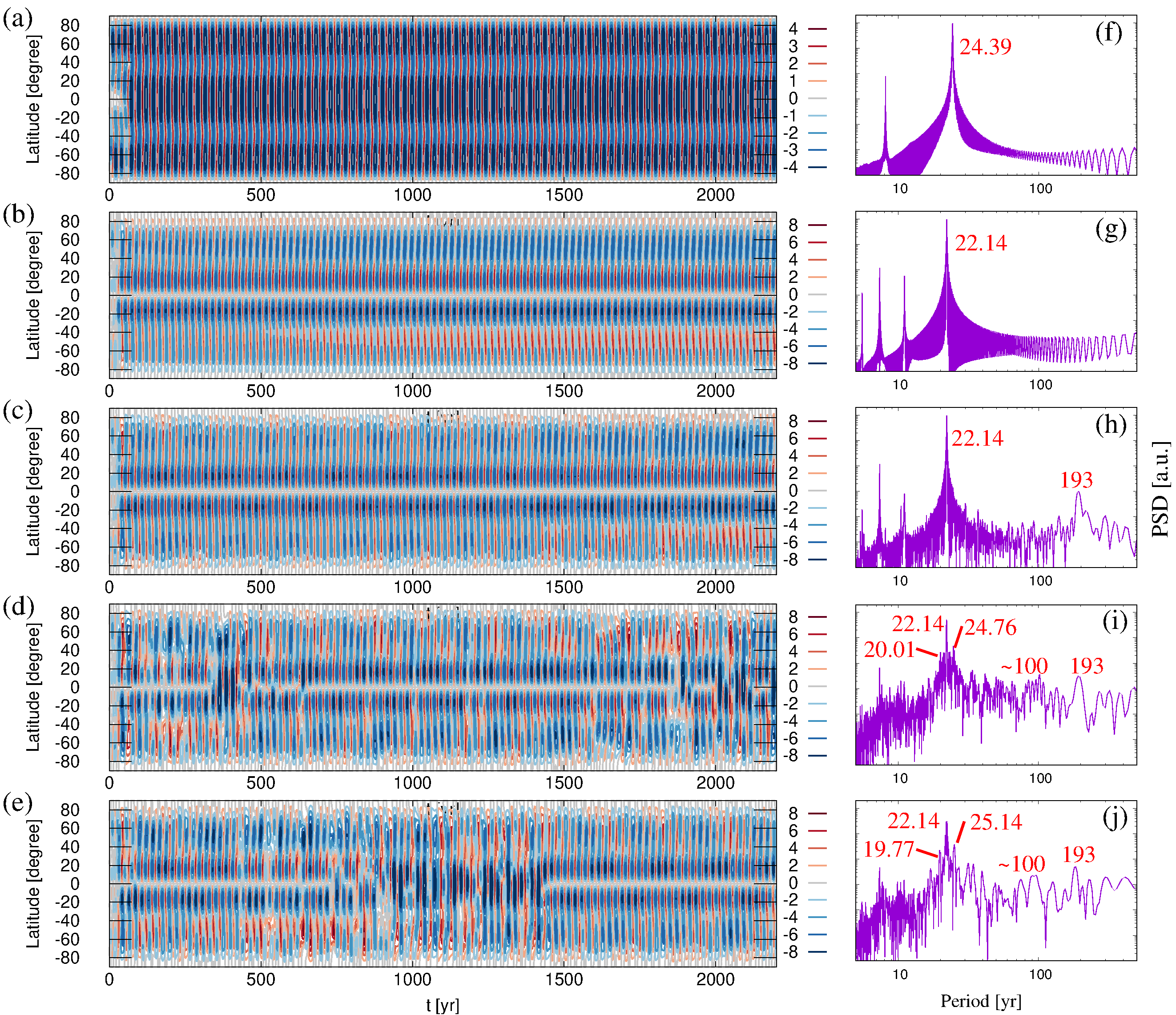}
  \caption{(a-e) Behaviour of $B(\theta,t)$, and (f-j) PSD  
  for $B(72^{\circ},t)$, with the common parameters
  $\omega_0=10000$, $\alpha^c_0=15$, $q^p_{\alpha}=0.2$, $q^c_{\alpha}=0.8$.
  The remaining parameters vary.
  (a,f) $\alpha^p_0=0$, $\kappa=0.5$, $D=0$: a  classical 
  $\alpha-\Omega$ dynamo 
  producing a {\it quadrupole} with 24.39-year period. 
  (b,g)  $\alpha^p_0=50$, $\kappa=0.75$, $D=0$: a tidally synchronized
  dynamo producing a {\it dipole} with 22.14-year period.
  (c,h) $\alpha^p_0=50$, $\kappa(t)=0.5+0.5 m(t)$, $D=0.2$:
  as (b,g), but with noise and 
  a modulation of $\kappa$ with an angular momentum function $m(t)$
  according to Fig.~3(a) (the maximum being normalized to 1).
  As seen in (h), this dipole solution contains a beat period of 193 years.
  (d,i) $\alpha^p_0=50$, $\kappa(t)=0.18+1.0 m(t)$, $D=0.215$:  similar to  
  (c,h), but with stronger $\kappa$ variation.
  (i) shows some peaks around 100 years, reminiscent of 
  the Gleissberg cycle(s). 
  (e,j) $\alpha^p_0=50$, $\kappa(t)=0.17+1.3 m_{\rm JS}(t)$, $D=0.23$; 
  similar to (d,i), but with a simpler angular momentum 
  $m_{\rm JS}(t)$ which is restricted to the 19.86-years periodic 
  part resulting from Jupiter and Saturn only. 
   }
  \label{fig:4}
\end{figure}

In the next step, see Fig.~4c, we assume a modulation of 
the parameter $\kappa$ with the angular 
momentum time series from Fig.~3 (plus some weak noise with $D=0.2$). 
Thereby, we obtain  a clear
additional peak (Fig.~4h) at the 193-year beat period 
between the underlying
22.14-year Hale cycle and the dominating (Jupiter-Saturn  related) 
19.86-year
period of the modulation of $\kappa$. Interestingly, this 
193-year signal is connected with the same type of variation of the 
``magnetic equator'' as observed in \cite{Pulkkinen1999} 
(although with a Gleissberg-type period in their case).
 
Figure 4d shows a similar dynamo run, this time with a stronger 
variation of
$\kappa$, which obviously leads to some intervening quadrupolar 
fields during the run.
The corresponding PSD (Fig.~4i) exhibits now also 
some Gleissberg-type peaks around 100
years. In parallel with that, the basic Hale cycle 
develops two side peaks
at around 20.0 years and 24.8 years, not dissimilar to 
the observed ones in Fig. 1c. Apparently, the dynamo undergoes an
intermediate 
locking close to the 19.86 years modulation cycle which, in turn, 
must be compensated by some longer cycles. It is these 
{\it prolonged} Hale cycles which
produce the {\it shortened} Gleissberg-type beat periods around 
100 years (see also Fig. 1c,d for plausibilization).

To clarify whether such Gleissberg-type peaks may alternatively
emerge as beat periods between the Schwabe cycle and the 
13.81-year Jupiter-Uranus and/or the 12.78-year Jupiter-Neptune
synode, we employ a simplified angular momentum time series 
where only the influence of Jupiter and Saturn is taken into account, 
whereas the frequencies due to other planets are omitted. 
The resulting PSD (Fig.~4j) continues to show some Gleissberg-type peaks, 
which - together with the sustained two side bands of the Hale cycle -  
speaks in favour of their non-linear origin discussed above. Yet,
the also observable reduction of the ''noisiness'' of the PSD 
(Fig.~4j compared
with Fig.~4i)
suggests at least some complementary role of the other frequencies 
which are present in the Sun's movement around the SSB. 

\section{Conclusions}
 
The Lomb-Scargle and wavelet 
analyses of Schove's solar cycle maxima data have reconfirmed
a Schwabe cycle with 11.07-year period, superposed by a clear Suess-de Vries cycle
with a period of appr. 200 years, and one or a few Gleissberg-type cycles around 
100-year periodicity. An extrapolation of the dominant harmonics points robustly 
to a next grand minimum in the 21st century, in accordance with previous 
predictions. The related decrease 
of solar activity may allow for a better
differential diagnostics of the respective weights of the 
two key climate drivers, solar activity/irradiance and anthropogenic 
greenhouse gases, whose individual effects were hard to disentangle 
in the course of their widely parallel rise during the last century.

Our tidally synchronized solar dynamo model, enhanced by a (yet poorly understood) 
spin-orbit coupling effect based on the dominant 19.86 years period of 
Jupiter-Saturn synodes, 
has lead to a clear spectral peak at the beat period of 193 years, which is close
to the observed Suess-de Vries cycle. 
The robustness of this cycle, in turn, lends 
also greater plausibility
to the clocked character of the underlying 22.14 Hale cycle 
(a simple random-walk process with average 22.14-years periodicity, 
but large phase-shifts, would hardly show such
a beat period).
Furthermore, for sufficiently strong modulation we observe  
two side-bands of the Hale cycle, one centered 
around the 19.86-year Jupiter-Saturn period, the other one 
around 24.5 years (to compensate for the
``too short'' cycles, when keeping pace with the basic 
11.07 years tidal forcing).
It appears that those prolonged Hale cycles can produce 
new Gleissberg-like 
beat periods around 100 years. However, a complementary 
explanation in terms of beat periods between the 
Schwabe cycle and the Jupiter-Uranus/Neptune synodes  
cannot be completely excluded.

What we have not aimed at in this paper is any 
prediction of the very timing of grand minima. 
This would require a good understanding of the 
phase relation between 
the two basic physical mechanisms 
underlying this work: tidal synchronization of helicity oscillations, and
coupling of the Sun's orbital motion around the SSB to internal motion.  
It also remains to be seen whether or not the Hallstatt cycle could be 
incorporated into the concept. 
The consistent reproduction of the Schwabe cycle, its two side-bands 
(flanking the Wilson gap), 
the Suess-de Vries and some Gleissberg-type cycles, as accomplished 
in this paper, might be a promising starting point for further investigations.

\section*{Acknowledgments}
This project has received funding
from the European Research Council (ERC) under the
European Union's Horizon 2020 research and innovation programme
(grant agreement No 787544).
It was also supported in frame of the Helmholtz - RSF
Joint Research Group ``Magnetohydrodynamic instabilities: Crucial
relevance for large scale liquid metal batteries and
the sun-climate connection'',
contract No HRSF-0044 and RSF-18-41-06201.
Inspiring and helpful discussions with
J\"urg Beer, Alfio Bonanno, Antonio Ferriz Mas, Peter Frick,
Laur\`{e}ne Jouve,
G\"unther R\"udiger, Dmitry Sokoloff, Willie Soon,
Steve Tobias, Ian Wilson, and Teimuraz Zaqarashvili are 
gratefully acknowledged.





\lastpageno	


\end{document}